\documentclass[review]{elsarticle}
\usepackage{csquotes}
\usepackage{color}
\usepackage{xcolor}
\usepackage{subcaption}
\usepackage{booktabs}
\usepackage{amsmath}
\usepackage{amssymb} 
\usepackage{graphicx}
\usepackage{mathtools}
\usepackage{float}
\usepackage{lineno,hyperref}
\modulolinenumbers[5]
\usepackage[toc,title,page]{appendix}
\usepackage{natbib}
\usepackage{graphicx} 
\graphicspath{ {./images/} }
\usepackage[rightcaption]{sidecap}
\usepackage{wrapfig}
\begin{document}

\begin{frontmatter}

\title{Numerical simulation of the coalescence-induced polymeric droplet jumping on superhydrophobic surfaces}
\author[mymainaddress]{Kazem Bazesefidpar\corref{mycorrespondingauthor}}
\cortext[mycorrespondingauthor]{Corresponding author}
\ead{kazemba@mech.kth.se}

\author[mymainaddress,mysecondaryaddress]{Luca Brandt}
\author[mymainaddress]{Outi Tammisola}
\address[mymainaddress]{SeRC (Swedish e-Science Research Centre) and FLOW, Dept. of Engineering Mechanics, KTH Royal Institute of Technology, SE-10044 Stockholm, Sweden}
\address[mysecondaryaddress]{Dept. of Energy and Process Engineering, Norwegian University of Science andTechnology (NTNU), Trondheim, Norway}

\begin{abstract}
Self-propelled jumping of two polymeric droplets on superhydrophobic surfaces is investigated by three-dimensional direct numerical simulations. Two identical droplets of a viscoelastic fluid slide, meet and coalesce on a surface with contact angle 180 degrees. The droplets are modelled by the Giesekus constitutive equation, introducing both viscoelasticity and a shear-thinning effects. The Cahn-Hilliard Phase-Field method is used to capture the droplet interface. The simulations capture the spontaneous coalescence and jumping of the droplets. The effect of elasticity and shear-thinning on the coalescence and jumping is investigated at capillary–inertial and viscous regimes. The results reveal that the elasticity of the droplet changes the known capillary–inertial velocity scaling of the Newtonian drops at large Ohnesorge numbers; the resulting viscoelastic droplet jumps from the surface at larger Ohnesorge numbers than a Newtonian drop, when elasticity gives rise to visible shape oscillations of the merged droplet. The numerical results show that polymer chains are stretched during the coalescence and prior to the departure of two drops, and the resulting elastic stresses at the interface induce the jumping of the liquid out of the surface. This study shows that viscoelasticity, typical of many biological and industrial applications, affects the droplet  behaviour on superhydrophobic and self-cleaning surfaces.
\end{abstract}


\end{frontmatter}

\section{Introduction}
When two droplets coalesce, the total surface area decreases. Hence, surface energy is released during this process. If the two droplets are far from a wall, the new bigger drop oscillates symmetrically until the released surface energy has been dissipated by viscosity. However, when two drops of micro- or nanometer size coalesce on a superhydrophobic surface, the presence of a repellent wall breaks the vertical symmetry and the resulting droplet propels in the direction perpendicular to the wall \citep{Bor2009}. Coalescence-induced jumping has been reported on a variety of natural repellent surfaces such as cicada, lacewings \citep{Wisdom2013} and gecko skin \citep{Watson2015}, and can be exploited in a variety of applications such as anti-icing \citep{Zhang2013} and self-cleaning surfaces \citep{Watson2015, Wisdom2013}, and to control heat transfer \citep{Enright2014}. Several researchers have studied the different aspects of the coalescence-induced droplet jumping numerically and experimentally, including the basic mechanism of the two equal-sized drop self-propelled jumping \citep{Liu2014}, and the effects of droplet size mismatch \citep{Wasserfall2017,Wang2019}, droplet initial velocity \citep{Chu2020,Li2020}, surface topology \citep{Wang2016, Peng2020, Yuan2019, Lo2014, Mulroe2017}, surrounding gas properties \citep{Farokhirad2015,Vahabi2017,Yan2019}, and surface wettability \citep{Cha2016}. A few main results are outlined in the following.
 
When two equal-sized static drops coalesce on a superhydrophobic surface, their total surface area decreases. This implies that surface energy is released and converted into viscous dissipation and kinetic energy, in a proportion determined by the Ohnesorge number, which represents the ratio between viscous and capillary-inertial forces.  At large Ohnesorge numbers, corresponding to the viscous regime, the kinetic energy is completely absorbed by viscous forces, preventing the jumping of the merged droplet \citep{Liu2014}. Even at small Ohnesorge numbers, corresponding to the capillary–inertial regime, only less than 4 \% of the released surface energy converts to vertical translational kinetic energy, which nevertheless causes the jumping of the merged droplet. The conversion rate of surface energy into kinetic energy reduces when the droplets are of unequal sizes, due to the strong asymmetric flow  \citep{Wasserfall2017,Wang2019}. The merged droplet attains an asymmetric shape and jumps with an oblique angle when one of the two droplets has an initial velocity; moreover, the jumping velocity of the merged droplet increases significantly above a critical initial velocity \citep{Li2020}. 

In addition, macrostructures on the surface affect the jumping velocity and energy transfer rates significantly. The jumping velocity and the conversion efficiency of surface energy to kinetic energy decrease if the lower contour of the merging drop falls between the gap of two rectangular grooves, whereas both jumping velocity and energy conversion increase when the liquid bridge expands on a triangular prism structure \citep{Wang2016}. The critical Ohnesorge number for droplet jumping depends on both the surface wettability and the ambient fluid properties such as density and viscosity; in particular, a larger density contrast between the ambient and drops will cause the merged droplet to jump higher \citep{Farokhirad2015}.

In very recent experiments, the effect of the drops' elasticity on the coalescence process was studied for both freely suspended drops and sessile drops with radius $O(1)$ micrometer on the hydrophobic surfaces \citep{Dekker}. They found that elasticity enhances the curvature of connecting bridge between two merging drops, and polymer stresses remain confined in a small region around the liquid bridge between the coalescing drops. However, the induced elastic stresses were found to be insufficient to alter the temporal evolution of the bridge in the capillary–inertial regime, and hence elasticity did not change the flow regime.

In the present work, we perform numerical simulations to study the effects of the non-Newtonian viscoelastic properties of two equal-sized static droplets on the coalescence-induced droplet jumping at large Ohnesorge numbers. Our studies extend from the viscous-capillary to the inertial-capillary regime, and in the former case we do observe prominent changes due to elasticity. We use the Cahn-Hilliard Phase-Field method for capturing the interface between the two phases, and the Giesekus constitutive equation to model the viscoelasticity of the drops. First, the role of elasticity is investigated by comparing the vertical velocity and different components of energy for a Newtonian and Oldroyd-B droplet at the same Ohnesorge number based on the same zero shear viscosity, while the influence of the liquid shear-thinning rheology is examined by using Giesekus model. 
\section{Governing equations and Numerical methods}
The numerical method used in this work has been described in detail in \citet{Bazesefidpar2021}, so we only give a brief outline here. We consider two immiscible fluids with different densities and viscosities. The outer fluid is Newtonian with viscosity $\mu_{n}$, whereas the droplets consist of a Giesekus fluid with solvent viscosity $\mu_{s}$, polymeric viscosity $\mu_{p}$, and the other non-Newtonian rheological properties as below. 
To distinguish between the phases, we introduce a phase-field variable, where $\phi = \pm 1$ in the bulk fluids and $\phi = 0$ at the fluid/fluid interface. This problem can be modelled with the following coupled equations \citep{Yue2004,Abels2012}:
\begin{eqnarray}
	\rho(\frac{\partial{\mathbf{u}}}{\partial{t}}+({\mathbf{u}}\cdot\nabla){\mathbf{u}})+
	{\mathbf{J}}\cdot\nabla{\mathbf{u}}=-\nabla{p}+\nabla\cdot{\boldsymbol\tau} 
	+\nabla\cdot\mu_{s}(\nabla{{\mathbf{u}}}+
	\nabla{{\mathbf{u}}^T})+G\nabla\phi,
	\label{NS2}
\end{eqnarray}
\begin{eqnarray}
	\nabla\cdot{\mathbf{u}}=0,
	\label{NS3}
\end{eqnarray}
the Cahn-Hilliard model:
\begin{eqnarray}
	\frac{\partial{\phi}}{\partial{t}}+\nabla\cdot({{\mathbf{u}}\phi})= 
	\nabla\cdot(M\nabla G),
	\label{NS1}
\end{eqnarray}
\begin{eqnarray}
	G=\lambda(-\nabla^2{\phi}+\frac{1}{\eta^2}\phi(\phi^2-1)), 
	\label{NS5}
\end{eqnarray}
and the Giesekus constitutive model:
\begin{eqnarray}
	\boldsymbol\tau_{p}+\lambda_H(\frac{\partial{\boldsymbol\tau_{p}}}{\partial{t}}+
	{\mathbf{u}}\cdot\nabla{\mathbf{\boldsymbol\tau_{p}}}-\boldsymbol\tau_{p}\nabla{{\mathbf{u}}}-
	\nabla{{\mathbf{u}}^T}\boldsymbol\tau_{p})+\frac{\alpha\lambda_H}{\mu_p}(\boldsymbol\tau_{p}\cdot\boldsymbol\tau_{p})&=& \nonumber \\
	\mu_p(\nabla{{\mathbf{u}}}+\nabla{{\mathbf{u}}^T}), \quad \quad \quad \quad \quad \quad \quad \quad \quad \quad \quad \quad \quad \quad \quad \quad \quad \quad \quad & &
	\label{NS4}
\end{eqnarray}

\begin{eqnarray}
	\boldsymbol\tau=\frac{(1+\phi)}{2}\boldsymbol\tau_{p}, 
	\label{NS6}
\end{eqnarray}
In the above equations, $\mathbf{u}(\mathbf{x},t)$ is the velocity vector, $p(\mathbf{x},t)$ is the pressure, and $\boldsymbol\tau(\mathbf{x},t)$ is the extra stress due to the polymers, equal to $\boldsymbol\tau_p$ inside the droplet and  $0$ outside (see eq. \ref{NS6}). In the Cahn-Hilliard equation, $G$ is the chemical potential, $M$ is the mobility parameter, and $\eta$ is the capillary width  of the interface. In the Giesekus model, $\boldsymbol\tau_{p}$ is the polymer stress, $\lambda_H$ is the polymer relaxation time, $\alpha$ is the Giesekus mobility parameter, and the polymeric retardation time can be related to the polymeric relaxation time by  $\lambda_{r}=\frac{\mu_{s}}{\mu_{s}+\mu_{p}}\lambda_{H}$. In Eq. (\ref{NS5}), $\lambda$ is the mixing energy density, and it is related to the surface tension in the sharp-interface limit \citep{Yue2004} by:
\begin{eqnarray}
	\sigma=\frac{2\sqrt{2}}{3}\frac{\lambda}{\eta}
	\label{NS7}
\end{eqnarray}
Fluid $1$ indicates the droplet phase and fluid $2$ represents the surrounding fluid (air). The density $\rho$ and the dynamic viscosity $\mu$ fields are expressed using the phase-field variable as:
\begin{eqnarray}
	\rho = \frac{(1+\phi)}{2}{\rho_1}+\frac{(1-\phi)}{2}{\rho_2},
	\label{NS8}
\end{eqnarray}
\begin{eqnarray}
	\mu=\frac{(1+\phi)}{2}{\mu_{s_1}}+\frac{(1-\phi)}{2}{\mu_{s_2}},
	\label{NS9}
\end{eqnarray}
The total viscosity of the non-Newtonian phase is $\mu_{t}=\mu_{s}+\mu_{p}$. The density satisfies the following relation\citep{Abels2012}\\
\begin{eqnarray}
\frac{\partial{\rho}}{\partial{t}}+\nabla\cdot{\rho\mathbf{u}}=-\nabla\cdot\mathbf{J},
\label{NSco}
\end{eqnarray}
where $\mathbf{J}=-\frac{(\rho_1-\rho_2)}{2}M\mathbf{\nabla}{G}$. Boundary conditions imposed on the substrate are, following \citet{Jacqmin2000,Qian2003}, the no-slip boundary condition for the velocities:
\begin{eqnarray}
	\boldsymbol{u}=\mathbf{0},
	\label{NS10}
\end{eqnarray}
and the static contact angle $\theta_s$ for the phase-field variable:
\begin{eqnarray}
	{\mathbf{n}}\cdot\nabla{\phi}+\frac{1}{\lambda}f_w^\prime(\phi)=0,
	\label{NS11}
\end{eqnarray}
\begin{eqnarray}
	f_{w}(\phi)=\sigma\cos(\theta_s)\frac{\phi(\phi^2-3)}{4}+\frac{(\sigma_{w_1}+\sigma_{w_2})}{2},
	\label{NS12}
\end{eqnarray}
where $\mathbf{n}$ is the outward pointing normal vector to the boundary, and $f_{w}(\phi)$ is a function describing the fluid–solid interfacial tension. \\
A second-order accurate scheme is employed for the temporal discretization of Eq. (\ref{NS1}) and (\ref{NS2}) while a semi-implicit splitting  scheme is used to treat the linear parts implicitly and the non-linear parts explicitly \citep{Dong2012}. To avoid the High-Weissenberg number problem (HWNP), the log-conformation reformulation (LCR) of equation Eq. (\ref{NS4}) \citep{Fattal2004,Fattal2005} is used and advanced in time by a second-order total variation diminishing (TVD) Runge-Kutta method \citep{Gottlieb1998}. Finally, we use second-order central differences to approximate spatial derivatives, except for the advection terms in Eq. (\ref{NS1}) and (\ref{NS4}), where the fifth-order WENO-Z is used to improve stability and accuracy \citep{Borges2008}

\begin{figure}[tbp]
	\centering
	\includegraphics[width=0.6\textwidth]{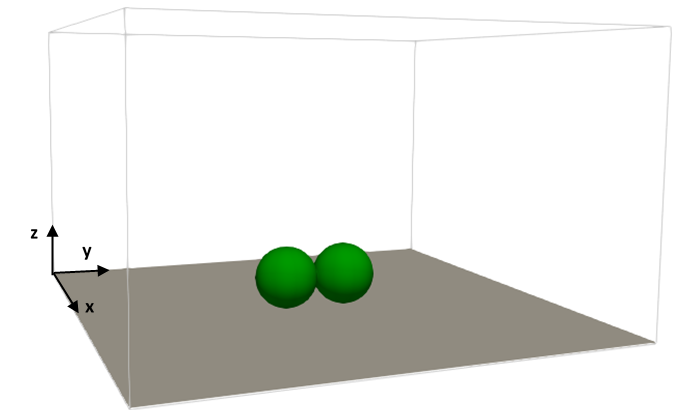}
	\caption{Sketch of the chosen computational domain $\Omega_1$.}
	\label{fig:fig1}
\end{figure}

\begin{figure}[tbp]
	\centering
	\includegraphics[width=0.55\textwidth]{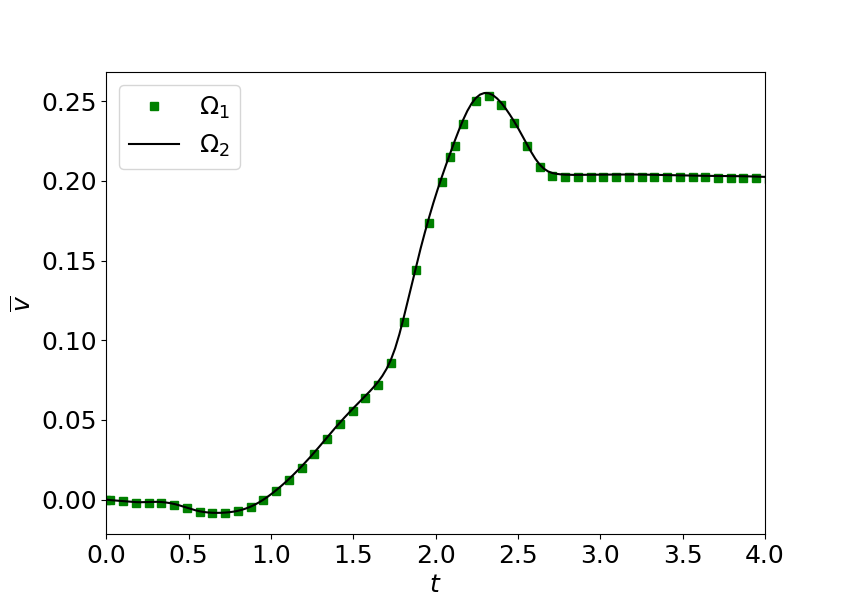}
	\caption{Average velocity integrated over the droplet volume in time, for the chosen domain (stars) and a bigger domain $\Omega_{2}$ (solid line), to show independence of the domain size. The parameters are $Oh=0.0076$ and $De=10$.}
	\label{fig:fig0}
\end{figure}

\begin{figure}[tbp]
	\centering
	\includegraphics[width=0.55\textwidth]{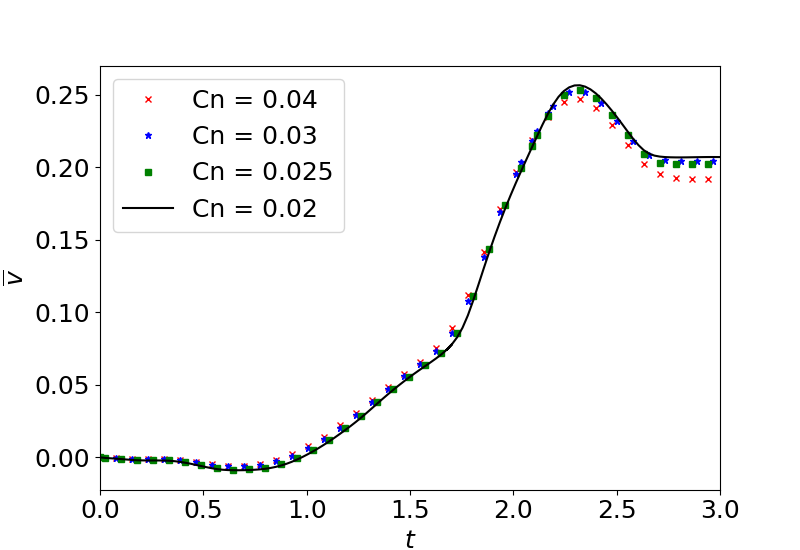}
	\caption{The evaluation of average velocity of the merged viscoelastic droplet for different $Cn$ numbers.}
	\label{fig:fig2}
\end{figure}
\section{Physical model and computational domain}
We consider two equal-sized initially static viscoelastic drops touching a homogeneous surface (Fig.\ \ref{fig:fig1}) with a static contact angle of $180^{o}$.  When two adjacent droplets coalesce on a superhydrophobic surface, the formed liquid bridge impinges on the substrate, and the merged droplet may jump above the substrate. Here, gravity is neglected, because the droplet radius is assumed to be much smaller than the capillary length and therefore capillary forces are expected to dominate. \\
The capillary-inertial velocity is chosen as the velocity scale \citep{Bor2009} $u_{ci}=\sqrt{\sigma /\left(\rho_{1}r_{0}\right)}$, and the droplets initial radius as the length scale. This gives rise to seven nondimensional numbers. Firstly, the Ohnesorge number $Oh=\left(\mu_{1}/\sqrt{\rho_{1}\sigma r_{0}}\right)$ representing the relative importance of viscous to capillary-inertial forces; the Weissenberg number $Wi=\left(\lambda_{H}u_{ci}/r_{0}\right)$ representing the ratio between elastic and viscous forces; the Peclet number $Pe=\left(2\sqrt{2}u_{ci}r_{0}\eta\right)/\left(3M\sigma\right)$ representing the ratio between the advection and diffusion in the Cahn-Hilliard equation. Furthermore, the Cahn number $Cn=\left(\eta/r_{0}\right)$ is the ratio between the interface width and the characteristic length scale; $\beta=\mu_{s}/\left(\mu_{s}+\mu_{p}\right)$ is the ratio between the polymeric viscosity and total viscosity; $k_{\mu}=\left(\mu_{2}/\mu_{1}\right)$ is the ratio between the ambient and droplet viscosities; $k_{\rho}=\left(\rho_{2}/\rho_{1}\right)$ between the ambient and droplet densities. The different components of the  energy are scaled by $\sigma r_{0}^{2}$. In what follows, all quantities will be nondimensional unless indicated otherwise.

To quantify the role of the fluid elasticity on the droplet jumping, we will measure the mass-averaged velocity of the droplet, defined as:
\begin{eqnarray}
\overline{v} = \frac{\int\limits_{\Omega} \frac{1}{2}(1+\phi) v_{z} \ d\Omega}{\int\limits_{\Omega}\ \frac{1}{2}(1+\phi) d\Omega}
\label{NS13}
\end{eqnarray}
where $\Omega$ is the computational domain (see Fig. \ref{fig:fig1}), and $z$ the direction perpendicular to the solid substrate. We also analyze the different components of the energy during the coalescence and jumping. The total energy $E_T$ of an Oldroyd-B fluid is the sum of the surface energy $E_s$, kinetic energy $E_k$, and elastic energy $E_e$, defined in phase-field framework as \citep{Boyaval2009,Mokbel2018}:

\begin{eqnarray}
	\begin{aligned}
		E_{s} &= \int\limits_{\Omega} \frac{3Cn}{4\sqrt{2}}\left[{\mid \nabla{\phi} \mid}^{2}+\frac{1}{2 Cn^{2}}{(\phi^{2}-1)}^{2}\right] \ d\Omega,\\
		E_{k} &= \int\limits_{\Omega} \frac{1}{2}\left(1+\phi\right)\mathbf{u}\cdot\mathbf{u} \ d\Omega,\\
		E_{e} &= \int\limits_{\Omega} \frac{Oh}{2}\frac{(1-\beta)}{Wi}\operatorname{tr}(\boldsymbol{c}-\ln{\boldsymbol{c}}-\boldsymbol{I}) \ d\Omega,\\
		E_{T} & = E_{s}+E_{k}+E_{e},\\
	\end{aligned}
	\label{NS14}
\end{eqnarray}
where the relationship between polymer stress $\mathbf{\boldsymbol\tau_{p}}$ and the conformation tensor $\mathbf{\boldsymbol{c}}$ is 
\begin{eqnarray}
	\mathbf{\boldsymbol\tau_{p}}=\frac{(1-\beta)}{Wi}\left(\mathbf{\boldsymbol{c}}-\mathbf{\boldsymbol{I}}\right),
	\label{NS15}
\end{eqnarray}
The part of the kinetic energy associated to the vertical velocity component is most relevant here as it can be associated with the jumping motion, while the rest is related to interface oscillatory motions \citep{Liu2014}. We will therefore consider a translational kinetic energy, defined as,
\begin{eqnarray}
	E_{k,tr} = \int\limits_{\Omega} \frac{1}{2}(1+\phi)\overline{v}^{2} \ d\Omega
	\label{NS16}
\end{eqnarray}
where $\overline{v}$ is the droplet mass-averaged velocity in z-direction, and $E_{k,os}=E_{k}-E_{k,tr}$ is the part of the kinetic energy associated with the oscillatory motion.

The numerical setup is as follows. The nondimensional domain size is chosen of size $\Omega=[0,10]\times[0,10]\times[0,8]$. Two adjacent droplets with initial radius $1$ are placed above the x--y plane at z=0, see the Fig. \ref{fig:fig1}. We impose no-slip and no-penetration conditions on the two boundaries in the z-direction, with static contact angles $\theta_{s}=180^{o}$ at the bottom wall and $\frac{\partial\phi}{\partial n}=0$ at the top boundary. 

Periodic boundary conditions are applied for all variables in the x- and y-directions. We use 8 grid points across the nominal interface in order to resolve the sharp gradients, and set the Peclet number to $Pe=\frac{6}{Cn}$, according to the guidelines in \citet{Magaletti2013,Xu2018} to approach the sharp-interface limit.

To ensure that the chosen grid and domain size are sufficient, we performed the following numerical tests.
Firstly, we examined the effect of the computational domain on the velocity of the merged droplet by performing an additional  simulation on a larger domain 
$\Omega_{2}=[0,15]\times[0,15]\times[0,9]$; the results obtained on $\Omega_{1}$ match those obtained on the larger domain $\Omega_{2}$, as shown in Fig.\ \ref{fig:fig0}. We also tested the grid dependency of the results by comparing the averaged velocity of the merged viscoelastic droplet for four different resolutions corresponding to different values of the Cahn number. For this test, the following values of the dimensionless numbers introduced above were used:
\begin{align*}
	Oh&=0.0076,  &  Wi&=10,  &  \beta&=0.1,  &  k_{\mu}&=0.017,   &  k_{\rho}&=0.00119.
\end{align*}
Fig. \ref{fig:fig2} shows that the averaged velocity of the merged drop with $Cn=0.025$ is almost the same as that obtained with the finer grid $Cn=0.02$. Higher values of $Cn$ displays non-negligible differences when the simulation time exceeds 2. Thus, we choose $Cn=0.025$, corresponding to a grid with $N_{x}\times N_{y}\times N_{z}=760\times760\times608$ grid points. This satisfactory convergence is achieved adopting the scaling between the Peclet number and Cn number suggested by \cite{Magaletti2013,Xu2018}.

\begin{table}[tbp]
	\begin{center}
		\caption{Experimental fluid properties by \citet{Yan2019} at $25^{o}C$}
		\label{Tabel 1}
		\begin{tabular}{c c c c c c c c c}
			\hline
			$r_{0}(\mu m)$ & $\rho_{1}(\frac{kg}{m^{3}})$ & $\mu_{1}(Pa s)$ & $\sigma(\frac{N}{m})$ & $k_{\rho}$ & $k_{\mu}$ & $\theta_{a}^{app}$ & $\theta_{r}^{app}$ & $\Delta\theta^{app}$\\
			\hline
			290 & 998.2  & 0.001 & 0.072 & $\frac{1}{839}$  & $\frac{1}{58.8}$ & $170.3^{o}$ & $167.7^{o}$ & $2.6^{o}$\\
			\hline
		\end{tabular}
	\end{center}
\end{table}

\begin{figure}[tbp]
	\centering
	\subfloat{\includegraphics[width = 3.5in]{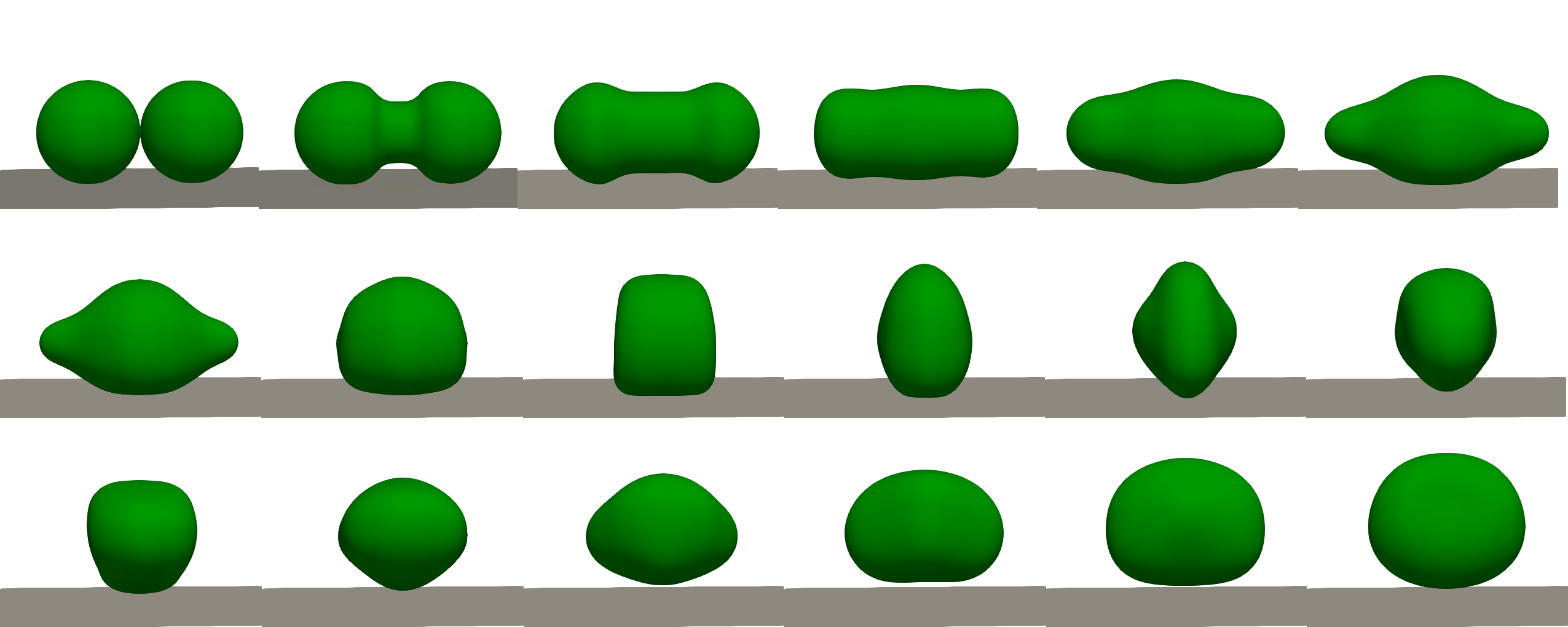}} \\
	\subfloat{\includegraphics[width = 3.5in]{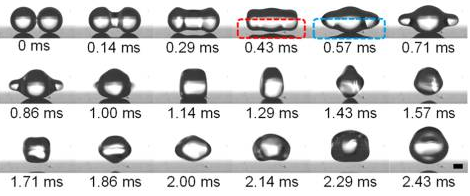}}
	\caption{The coalescence and jumping of two Newtonian droplets on a superhydrophobic surface at $Oh=0.0076$: (a) Numerical results from present work ($yz$ view) (b) Experimental data of \citet{Yan2019}.}
	\label{fig:fig3}
\end{figure}

\begin{figure}[t]
	\centering
	\hspace*{0.8cm}\subfloat[]{\includegraphics[width = 2.5in]{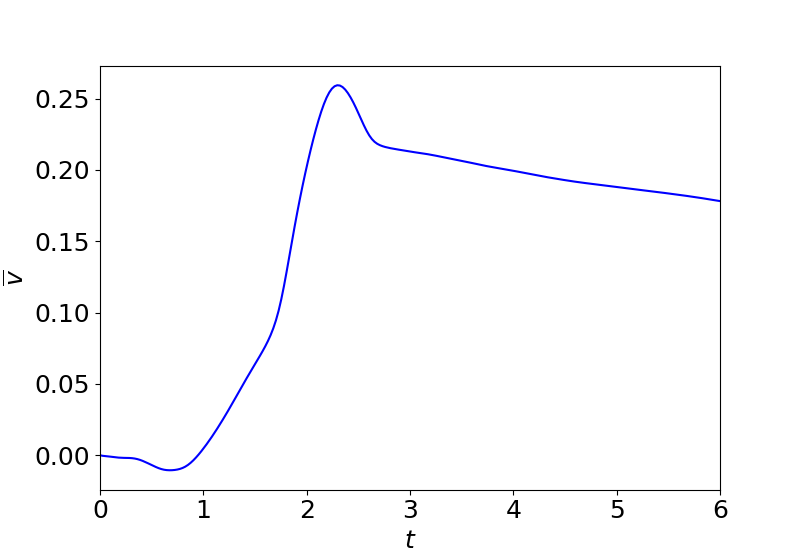}}\\
	\subfloat[]{\includegraphics[width = 2.5in]{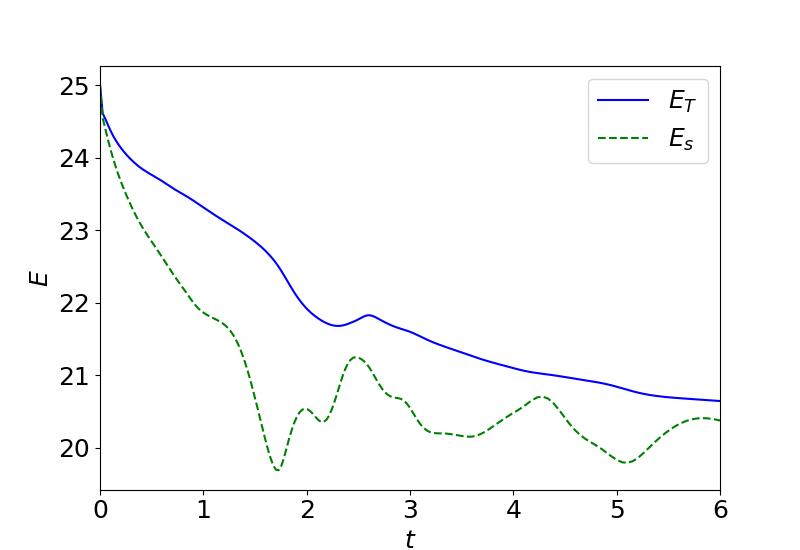}} 
	\subfloat[]{\includegraphics[width = 2.5in]{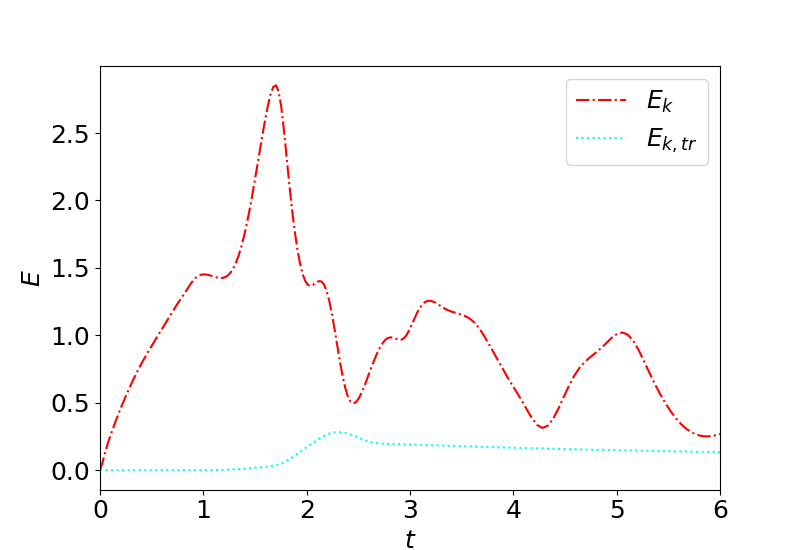}}
	\caption{Time-dependent quantities of the Newtonian merged drop on a superhydrophobic surface at $Oh=0.0076$: (a) The average velocity, (b) Total and surface energies of the merged drop (c) Kinetic and translational kinetic energies of the merged drop.}
	\label{fig:fig4}
\end{figure}

\section{Results}
For the results presented here, the density ratio and viscosity ratio are kept constant ro $k_{\rho}=\frac{1}{839}$ and $k_{\mu}=\frac{1}{58.8}$., following the values from the experiments in \citet{Yan2019}.
\subsection{Newtonian droplets - comparison with experiments}
The solver has been validated against several Newtonian and viscoelastic two-phase flow benchmarks in 2D and 3D \citep{Bazesefidpar2021}. Here, we compare the spontaneous coalescence and jumping motion of a Newtonian drop on a superhydrophobic surface with the experimental data of \citet{Yan2019}. We choose the same physical parameters as in the experiment, see Table \ref{Tabel 1}. The influence of the solid-liquid adhesion on the self-propelled jumping is negligible when the contact angle hysteresis ($\Delta\theta^{app}$) is less than $10^{o}$ \citep{Vahabi2017,Xu2016}, so we ignore the contact angle hysteresis and impose a static contact angle $\theta_{s}=180^{o}$ on the bottom wall. Fig. \ref{fig:fig3} presents the experimental data \citep{Yan2019} for the coalescence of two Newtonian drops on a superhydrophobic surface and the corresponding numerical results; the visualization shows that the numerical simulation is able to capture the coalescence and jumping process accurately in time.

The jumping velocity of the merged water (Newtonian) drop on a superhydrophobic surface is constant in the capillary-inertial region (i.e., $Oh\lesssim 0.1$); \cite{Bor2009} reported ${{v}_{j}}\approx0.2$ for the water drop at $19^{o}C$ on a textured superhydrophobic surface. Later, \cite{Yan2019} 
reduced the level of undesired external disturbances and measured a velocity ${{v}_{j}}\approx0.26$ for self-propelled jumping of water drops upon coalescence on a superhydrophobic surface.
Fig. \ref{fig:fig4}(a) shows the averaged velocity of the merged droplet in our simulation. The jumping velocity is measured from the time the bottom of the merged drop leaves the surface. \citet{Liu2014} suggested that a sensible time for extracting the jumping velocity is the first pseudo-equilibrium, which corresponds to the time when the axial lengths of the merged droplet in x and y-directions become equal. These axial lengths are measured with respect to an axis attached to the center of mass of the merged drop, and density ratio in their numerical simulation was $k_{\rho}=0.02$.  
Following this criterion, the jumping velocity of the Newtonian drop at $Oh=0.0076$ is ${{v}_{j}}\approx0.21$ in our simulations. It should be noted that the criterion of \citet{Liu2014} may work less well at our high density ratios, since the merged droplet's average velocity decreases very rapidly after its maximum, and this may lead us to underestimate the jumping velocity.
The dimensionless time corresponding to the first pseudo-equilibrium is $t\approx3.32$ (dimensional time $t^*\approx1.92 ms$) with the droplet leaving the surface at $t\approx2.71$ ($t^*\approx1.57 ms$) in both simulation and experiment. The averaged velocity at $t=2.71$ is ${\overline{v}}\approx0.23$, and the maximum averaged velocity is ${\overline{v}}_{max}\approx0.26$, see Fig. \ref{fig:fig4} (a). 

Fig. \ref{fig:fig4} (b) presents the time-evolution of the total and surface energy of the merging and jumping droplet. The total energy of the droplet decreases over time due to the viscous dissipation, and this decrease is larger in the merging process and prior to jumping, because there are highly localized velocity gradients around the liquid bridge during its impingement on the substrate. Fig. \ref{fig:fig4} (c) depicts the total and translational kinetic energy of the merged droplet, confirming that a small fraction of the released surface energy is converted into transitional kinetic associated to the jumping motion.

\subsection{Viscoelastic droplets - elasticity effect} 
Most of the existing studies are restricted to experiments with water droplets and numerical simulation of Newtonian drops; \cite{Yan2019} newly investigated the effect of the liquid internal hydrodynamics by conducting experiment for the self-propelled jumping upon coalescence on a superhydrophobic surface with ethanol-water and ethylene-glycol solutions. Their experiment shows that the properties of the droplet affect the coalescence and jumping process significantly. A very recent experimental study, however, addressed the effect of the drops' elasticity on the coalescence process \citep{Dekker}, and their findings will be referred to later in this section.
In the following, we investigate numerically how the droplet elasticity (Weissenberg number) influences the jumping process for different values of the Ohnesorge number. One way to vary the Ohnesorge number in experiments is to keep the physical properties constant and vary its radius, and we adopt this approach in our simulation. It should be noted that other parameters change also with the droplet radius; the parameters for each case are summarized in Appendix A.

\begin{figure}[H]
	\centering
	\hspace*{0.8cm}\subfloat[]{\includegraphics[width = 2.5in]{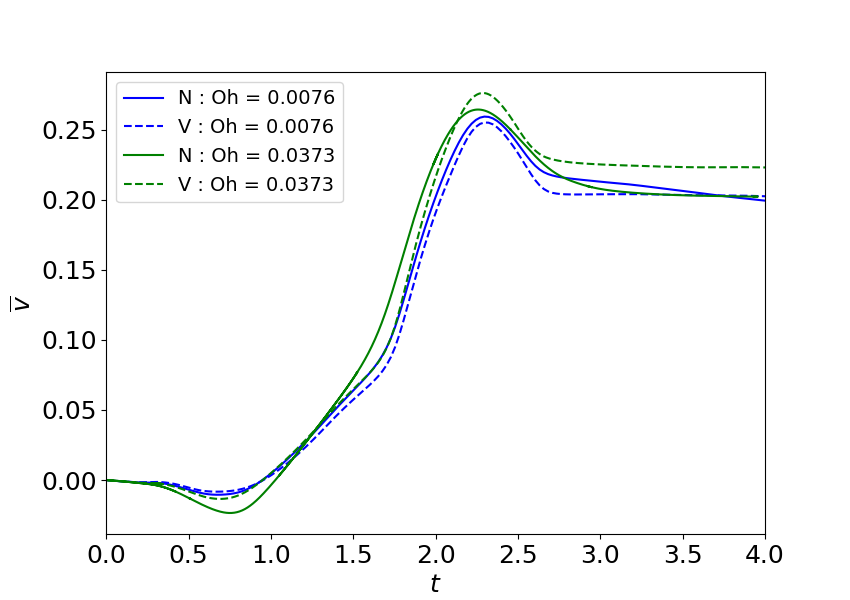}}\\
	\subfloat[]{\includegraphics[width = 2.5in]{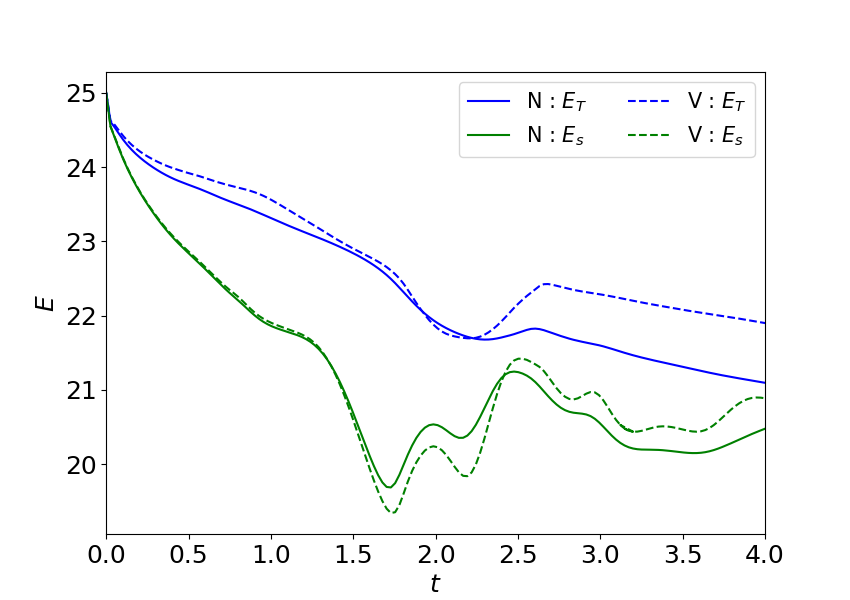}} 
	\subfloat[]{\includegraphics[width = 2.5in]{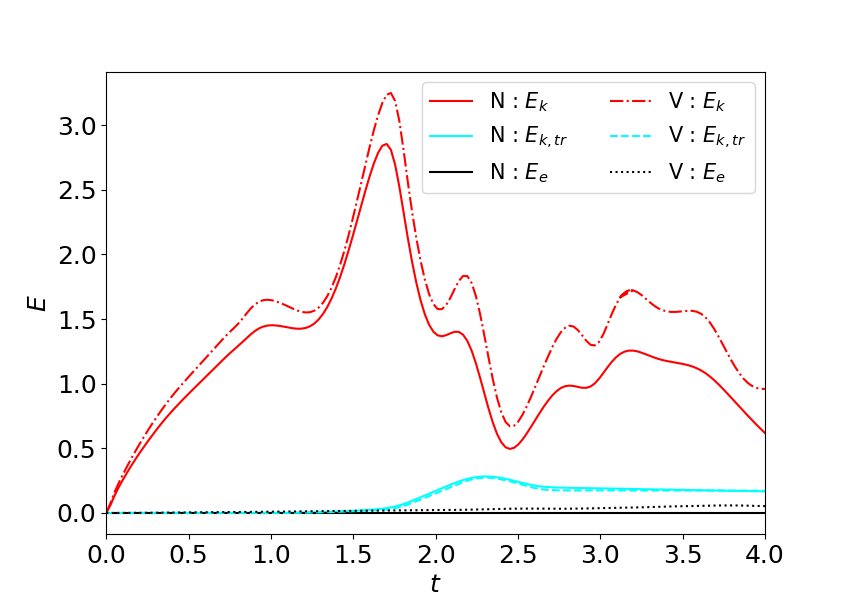}}
	\caption{Quantities of the merged Newtonian and Oldroyd-B drops at $Wi=10$ and $\beta=0.1$. N and V refer to Newtonian and Oldroyd-B drops respectively. (a) The averaged velocity (b) Total and surface energies of the merged drop at $Oh=0.0076$ (c) Kinetic and transitional kinetic energies of the merged drop at $Oh=0.0076$.}
	\label{fig:fig5}
\end{figure}
\subsubsection{Small Ohnesorge numbers}
To isolate the effect of the droplet elasticity on the coalescence and jumping on a superhydrophobic surface in the inertial-capillary region, $Oh\lesssim0.1$, we set $\alpha=0$ in the following simulations. This choice implies that shear-thinning is eliminated, and the Giesekus model reduces to the Oldroyd-B model.
The average velocities of the merged Newtonian and viscoelastic droplets are compared at two Ohnesorge numbers, $Oh=0.0076$ and $Oh=0.0373$. For Newtonian droplets, these values represent the capillary-inertial regime. The two dimensionless numbers defining the Oldroyd-B model are kept constant, $Wi=10$ and $\beta=0.1$, while the others vary since we are changing the droplet radius, see Appendix A.

Fig. \ref{fig:fig5} (a) shows the time evolution of the average velocity of the merged droplet for the Newtonian and viscoelastic cases, for both Ohnesorge numbers. The first observation is that both droplets jump from the surface at these low Ohnesorge numbers. Let us now consider the blue lines, corresponding to the smallest Ohnesorge number: Newtonian (solid line) and viscoelastic (dashed line). We observe that the elasticity of the drop has a negligible effect on the average velocity prior to and during jumping. However, there is a small qualitative difference after jumping, where the averaged velocity remains approximately constant for the viscoelastic droplet, while the velocity of the Newtonian droplet decreases in time. For the larger Ohnesorge number (green lines), we observe that the droplet elasticity increases the maximum averaged velocity. The total energy is dissipated more rapidly in the Newtonian droplet after departure, while there is less dissipation in the merged viscoelastic droplet, see Fig.\ \ref{fig:fig5} (b). The droplet kinetic energy is presented in Fig.\ \ref{fig:fig5} (c): the viscoelastic droplet has more kinetic energy so that it undergoes larger shape oscillations than the Newtonian droplet. The oscillations of the viscoelastic droplet are due to its elasticity and independent of the surface tension, see  \cite{Khismatullin2001}. The viscoelastic droplet oscillates even at the large Ohnesorge numbers corresponding to a highly viscous drop \citep{Khismatullin2001}.An extensional flow occurs when the two viscoelastic drops are coalescing, so polymer chains stretch and store elastic energy during the coalescence process, see Fig.\ \ref{fig:fig5} (c).

Summarizing, the average and jumping velocity are not considerably affected by the elasticity of the drops in the inertial-capillary regime. This result is in line with the experiments of \citet{Dekker}, where elasticity did not considerably influence the coalescence process in the inertial-capillary regime.  However, quantitatively we found that the oscillations are promoted by elasticity, and that the average velocity decays less rapidly after the droplet departure from the superhydrophobic surface.

\begin{figure}[t]
	\centering
	\hspace*{0.8cm}\subfloat[]{\includegraphics[width = 2.5in]{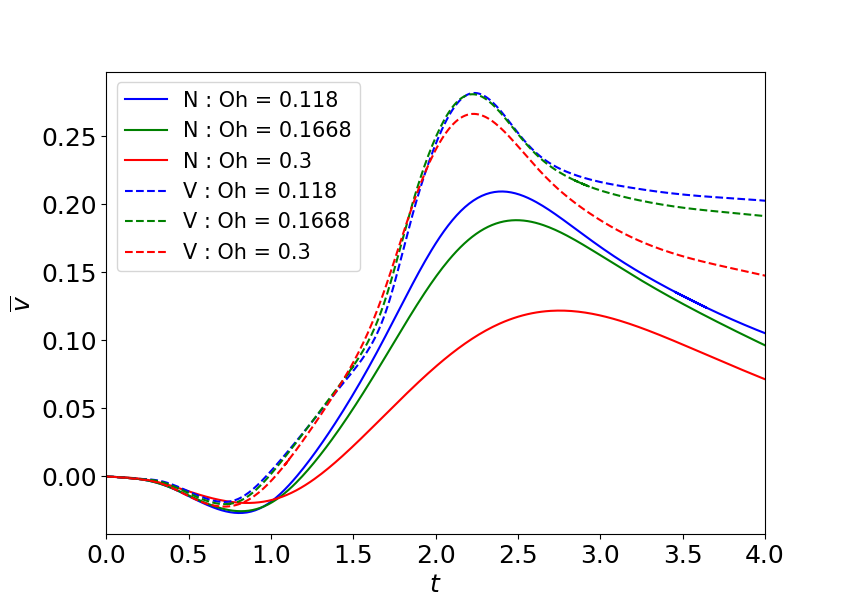}}\\
	\subfloat[]{\includegraphics[width = 2.5in]{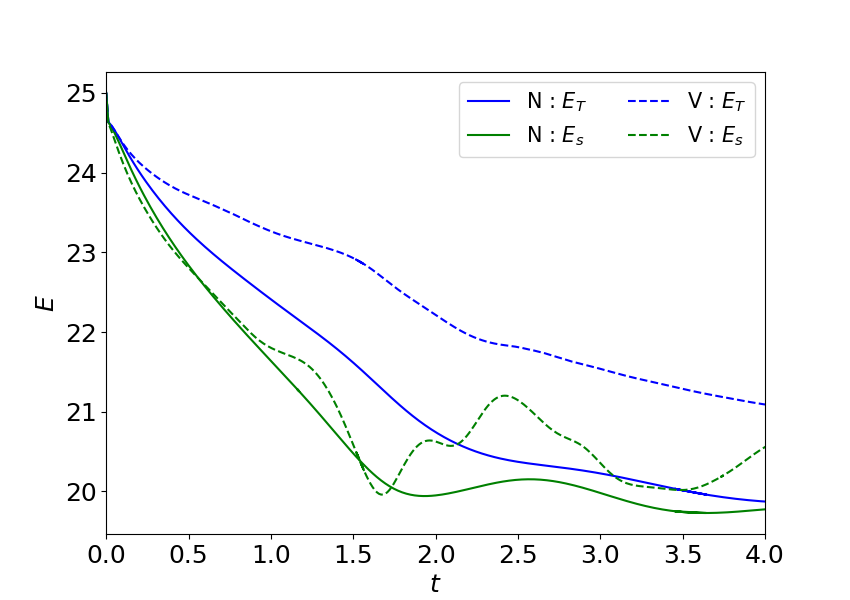}} 
	\subfloat[]{\includegraphics[width = 2.5in]{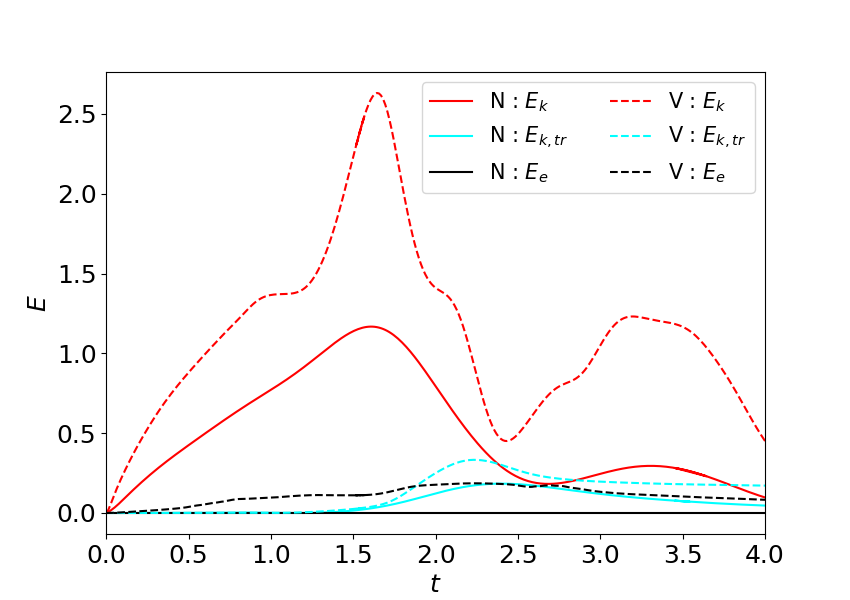}}
	\caption{Quantities of the merged Newtonian and Oldroyd-B drops at $Wi=10$ and $\beta=0.1$. N and V refer to Newtonian and Oldroyd-B drops respectively. (a) The averaged velocity (b) Total and surface energies of the merged drop at $Oh=0.118$ (c) Kinetic and transitional kinetic energies of the merged drop at $Oh=0.118$.}
	\label{fig:fig6}
\end{figure}

\begin{figure}[tbp]
	\centering
	\subfloat[]{\includegraphics[width = \textwidth]{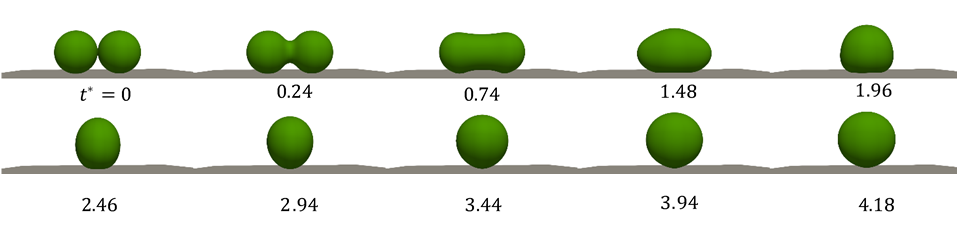}} \\
	\subfloat[]{\includegraphics[width = \textwidth]{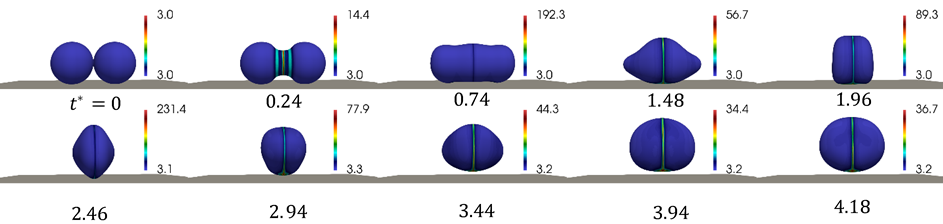}}
	\caption{The coalescence and jumping of Newtonian and Oldroyd-B drops on a superhydrophobic surface at $Oh=0.118$, $Wi=10$ and $\beta=0.1$ (a) Newtonian drops (b) Oldroyd-B drops. The trace of the dimensionless polymeric stresses are visualized on the surface of the polymeric drop.}
	\label{fig:fig7}
\end{figure}

\subsubsection{Large Ohnesorge numbers}
For Newtonian drops, viscous forces become the dominant at larger Ohnesorge numbers ($Oh\gtrsim0.1$), known as the viscous regime: both jumping and averaged velocities decrease rapidly with increasing Ohnesorge number due to the strong viscous dissipation. Let us now examine whether and how this behaviour changes for viscoelastic droplets at $Oh\gtrsim0.1$. 

The results from the simulations are reported in  Fig.\ \ref{fig:fig6}. First, we note that
the merged viscoelastic droplet gains much larger average velocity than the Newtonian one during the coalescence process, as seen by comparing the solid (Newtonian) and dashed (viscoelastic) lines of the same color in panel (a). Panel (b) of the same figure shows that the released energy is soon damped at $Oh=0.118$ ( ${t}^{*}\approx4$), and the total energy (blue solid line) reaches its equilibrium value $E_{T}^{*}=2^{\frac{2}{3}}4\pi$ consisting of surface energy only (green solid line). The Newtonian drop resulting from the coalescence reaches a spherical shape corresponding to its equilibrium, and stays on the surface without jumping, as shown in Fig. \ref{fig:fig7}. 

Let us now consider the polymeric drop (dashed blue and green lines). This merged polymeric drop has energy available to oscillate and move upwards until $t=4.0$. The kinetic energies of the Newtonian (solid) and Oldroyd-B droplets (dashed) are depicted in Fig.\ \ref{fig:fig6} (c), and both total kinetic and translational kinetic energies of the polymeric drop are larger than for the Newtonian drop. The Oldroyd-B drop also has the additional elastic energy due to the presence of the polymer molecules, see the black dashed line. These are stretched during the coalescence and prior to jumping, so that extra elastic energy is stored and available for the polymeric drop.

A visualisation of the coalescence and jumping of both droplets at $Oh=0.118$ is provided in Fig. \ref{fig:fig7}. The polymeric drops merge faster than the Newtonian drops, and the the bridge formed due to the coalescence reaches the substrate sooner. The interface (given by $\phi=0$) of the Oldroyd-B drops is coloured using the trace of the conformation tensor which indicates the intensity of polymer stretching.
As reported in experiments of \cite{Dekker}, the polymeric stresses are seen to be very concentrated around the merging interface and the capillary bridge. At the present flow regime, however, we observe significant changes due to elasticity in both the liquid bridge formation and in the merging and coalescence process. The merged polymeric drop undergoes a large deformation in all three directions and jumps out of the surface; this oscillatory motion and jumping is characteristic of the inertial-capillary region and is maintained at high $Oh$ when elasticity is present. Indeed, the  Newtonian drop goes rapidly towards its equilibrium condition without noticeable oscillations, and remains on the surface, as expected in the viscous region.

\begin{figure}[H]
	\centering
	\subfloat[]{\includegraphics[width = 3in]{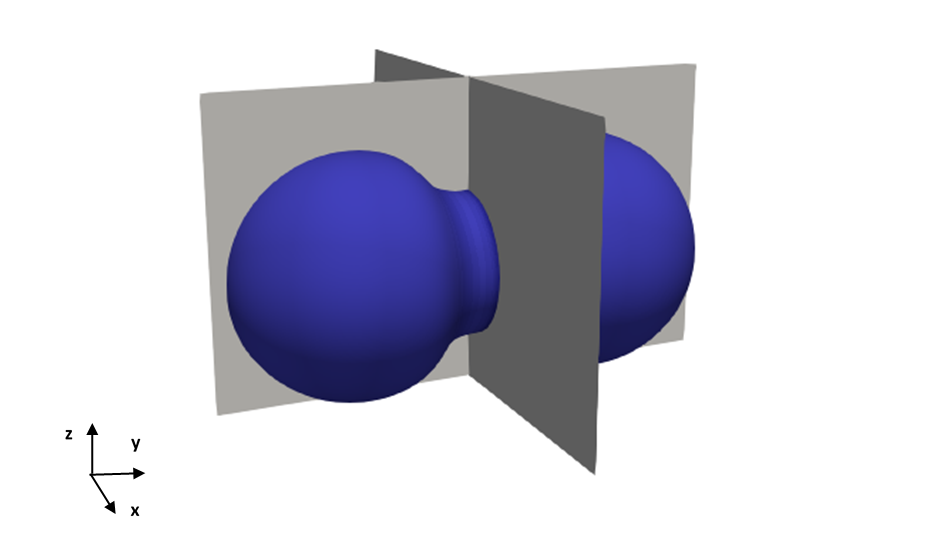}} \\
	\vspace*{0.5cm}
	\subfloat[]{\includegraphics[width=1\textwidth]{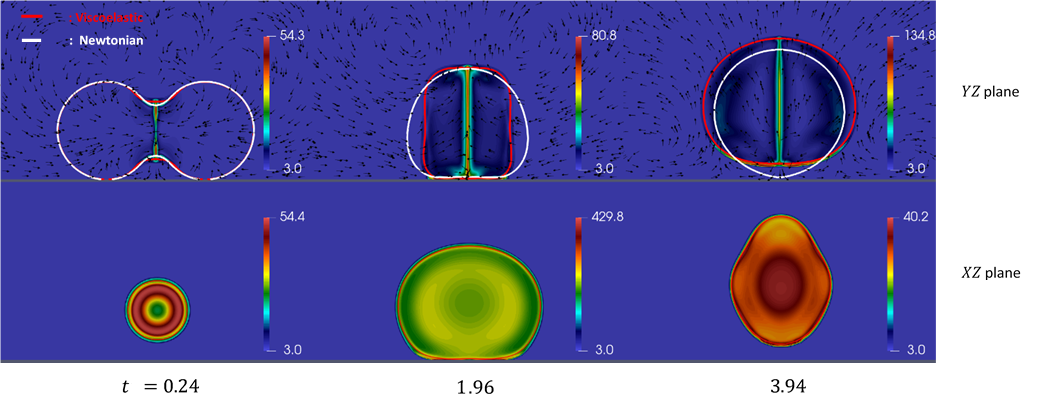}}
	\caption{
		Time evolution of viscoelastic drops coalescence on $XZ$ and $YZ$ planes on a superhydrophobic surface at $Oh=0.118$, $Wi=10$ and $\beta=0.1$. The interface $\phi=0$ of the merging Newtonian drops are depicted at $Oh=0.118$ on $YZ$ plane for the comparison : (a) $XZ$ and $YZ$ planes used for the visualization of the flow field (b) The interface of both Newtonian and viscoelastic drops on $YZ$ plane at three different times $t$ along with the velocity field and trace of conformation tensor belong to the viscoelastic drops. The trace of conformation tensor is visualized on $XZ$ for the viscoelastic drops.}
	\label{fig:fig8}
\end{figure}

To gain a better understanding on the effect of elasticity on the self-propelled jumping, the flow field is visualized on the central $XZ$ and $YZ$ planes inside the merging drops, see Fig.\ \ref{fig:fig8} (a). As concern the $YZ$ plane, shown in the top row, we depict the interface, identified by the $\phi=0$ value of the order parameter (red contour), velocity field (arrows), and trace of conformation tensor (colormap) for merging viscoelastic drop at three times, together with the the interface of merging Newtonian drop (white contour). This illustrates how the shape of Newtonian and viscoelastic droplets differ, and confirms the localisation of polymeric stresses at the merging cross-section. Moreover, the bottom row of the same figure displays contoues of the trace of the conformation tensor for the viscoelastic drops on the $XZ$ plane at the same dimensionless times as in Fig. \ref{fig:fig8} (b). The data indicate that polymers are most elongated prior to jumping, near the bottom wall.

A possible physical explanation for the polymer effect can be as follows. When the two initially static drops start to merge, the liquid  moves driven by the capillary pressure towards the center of the expanding bridge. Then, due to the conservation of mass, the liquid is forced to move in the transverse $XZ$ plane, see the velocity field of the merging viscoelastic drops in Fig.\ \ref{fig:fig8} (b) at $t=0.24$. 
This flow causes the polymer molecules to stretch in the $XZ$ plane and produce extra elastic stresses, which push the liquid bridge connecting the two polymeric drops to move faster. When the liquid bridge interacts with the substrate at $t\approx1$, the liquid is induced to move upwards due to the impermeability of the surface. This upward flow converges toward the $XZ$ plane and causes the polymers to stretch mainly in the vicinity of the substrate, as shown by the trace of conformation tensor at $t=1.96$ in Fig.\ \ref{fig:fig8} (b). 
The stretched polymers exert extra elastic stresses on the interface near the substrate, so the merged polymeric drop jumps out of the surface. The newly-formed larger drop leaves the surface at $t\gtrsim3$, when the trace of the conformation tensor decreases. 
Later, the polymers are mainly stretched at the bottom of the merged drop, and two small vortices appear in that region so that the polymer molecules remain stretched. Thus, these extra polymer stresses at the bottom of the drop push the drop to move upward. The polymers are also significantly stretched in the $XZ$ plane around the interface, and these extra polymer stresses push the interface to oscillate in the x-direction, see Fig. \ref{fig:fig8} (b).

These results reveal that the elasticity of the drop plays an important role at large Ohnesorge numbers in the coalescence and jumping processes of two initially static equal-sized polymer drops on a superhydrophobic surface; elasticity also affects the merged droplet motion after its departure as demonstrated by the oscillatory motion in highly-viscous yet viscoelastic drops.
\begin{figure}[htp]
	\centering
	\subfloat{\includegraphics[width = 3.5in]{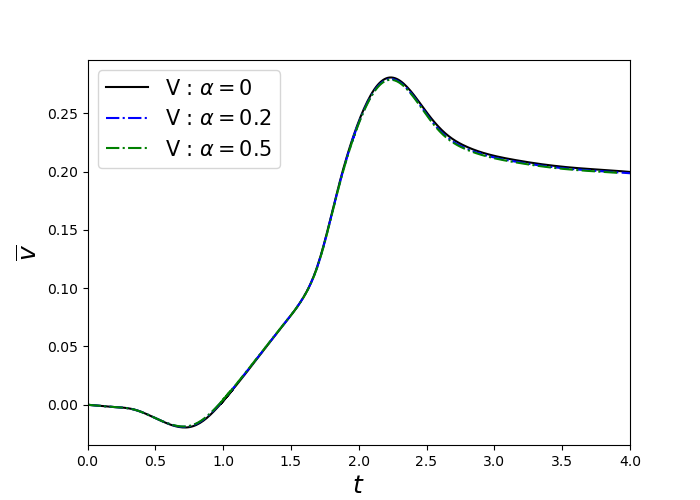}}
	\caption{The effect of drop shear-thinning $\alpha$ on the averaged velocity at $Oh=0.118$, $Wi=10$, and $\beta=0.1$.}
	\label{fig:fig9}
\end{figure}

\subsection{Viscoelastic droplets - effects of polymeric viscosity ratio and shear-thinning}
The effect of the shear thinning on the self-propelled jumping of two equal-sized polymeric drops has been studied by performing simulations at $Oh=0.118$, $De=10$, $\beta=0.1$ and varying $\alpha$; the results show that the effect of shear thinning is minor and negligible for the self-propelled jumping of two equal-sized polymeric drops even at large Ohnesorge numbers, see Fig.\ \ref{fig:fig9}.

\begin{figure}[htp]
	\centering
	\subfloat{\includegraphics[width = 3.5in]{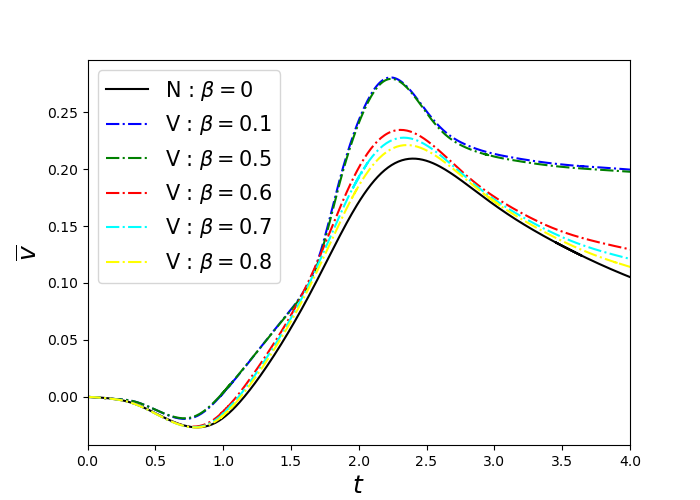}}
	\caption{The effect of polymeric viscosity ratio $\beta$ on the averaged velocity at $Oh=0.118$ and $Wi=10$.}
	\label{fig:fig10}
\end{figure}

In addition, we have investigated the effect of the polymeric viscosity ratio $\beta$ on the self-propelled jumping at $Oh=0.118$ and $Wi=10$. Fig.\ \ref{fig:fig10} depicts the variation of the averaged velocity of the droplet for $\beta=0.1-0.8$. The Newtonian droplet velocity is also shown for comparison. Two regimes can be distinguished: as long as $\beta\lesssim0.5$, we note a minor influence on the averaged velocity;  for $\beta\gtrsim0.6$, conversely, the velocity rapidly converges towards the Newtonian one, so that elasticity effects become negligible. 
This can be explained by considering the retardation time, \textit{i.e.} the relative time it takes for polymer molecules to be stretched. The retardation time can be related to the polymeric viscosity ratio by $\lambda_{r}=\beta\lambda_{H}$, while the flow time scale time is constant in our simulations since the droplet radius $r_{0}$, velocity scale $u_{ref}=\sqrt{\frac{\sigma}{\rho_{1}r_{0}}}$, and Weissenberg number $Wi$ are kept constant. Since the retardation time is increasing by increasing $\beta$, the polymers do not have time to stretch and store elastic energy during the coalescence and jumping when $\beta \gtrsim0.6$. Thus, the polymeric drops behave like Newtonian drops at large polymeric viscosity ratios.

\section{Conclusions and outlook}
In the present study, three-dimensional direct numerical simulations have been performed to study the self-propelled jumping of two equal-sized polymeric drops on a superhydrophobic surface with contact angle of $180^{o}$. The results demonstrate that the viscoelastic properties of the droplets have a significant impact on the coalescence and jumping. 

At small Ohnesorge numbers (inertial-capillary region), the elasticity effect is weak before the jumping; however, the averaged velocity of the coalesced drop does not decay as rapidly as for a Newtonian liquid. Drop shape oscillations are promoted bythe presence of the polymers.

At large Ohnesorge numbers ($Oh\gtrsim0.1$) however, profound differences between polymeric and Newtonian drops are observed during the coalescence and jumping process. The polymeric drops merge faster than the Newtonian drops, and the merged drop jumps out of the surface in contrast to their Newtonian counterparts, which remain on the substrate due to the large viscous dissipation. Our investigation reveals that the polymers are highly stretched at the cross-section of the merging droplets during coalescence, and these stretched chains exert extra elastic stresses on the interface of the merging drops in the vicinity of the wall, hence helping the polymeric drop to jump from the surface. These results are obtained with a typical value in the literature for the polymeric viscosity ratio, i.e.\ $\beta=0.1$; here, we also observe that the merged viscoelastic drop behaves like a Newtonian drop when $\beta\gtrsim0.7$. The larger $\beta$ corresponds to larger retardation times in our simulation, so that the polymer molecules do not have enough time to stretch. Finally, the shear-thinning effect is found to be negligible in the coalescence and jumping process of two equal-sized drops on a superhydrophobic surface.

Our results indicate that the elasticity of the droplet can change the viscous cutoff radius (for example $30\mu m$ for water) for the self-propelled jumping of drops on superhydrophobic surfaces. Thus, it is expected that polymeric drops jump from a superhydrophobic surface upon their coalescence with radii below the viscous cutoff radius for Newtonian drops at the same Ohnesorge number. Moreover, the merged polymeric drop oscillatory motion is promoted by the elasticity of the drop in both inertial-capillary and viscous-capillary regimes.

In this study, we have neglected the contact angle hysteresis, assuming it to be smaller than $10^{o}$; however, superhydrophobic surfaces may have large contact angle hysteresis, which might play an important role in the case of polymeric drops. Studying the effect of the contact angle hysteresis is one of the possible extensions of this work.

\section{Acknowledgments}
This work was funded by European Research Council (ERC) through Starting grant no. 852529 MUCUS, and by the Swedish Research Council through grant VR 2017-0489. We acknowledge the computing time on the supercomputer Beskow at the PDC center, KTH provided by SNIC (Swedish National Infrastructure for Computing),Sweden. 

\appendix
\section{}

In this Appendix, we report the values of the non-dimensional numbers used in the simulations.\\

\begin{table}[H]
	\begin{center}
		\caption{Dimensionless numbers used in the simulations at small Ohnesorge numbers, see section 4.2.1}
		\label{Tabel 2}
		\begin{tabular}{c c c c c c c c c}
			\hline
			$Case$ & Droplets & $Oh$ & $Wi$ & $\beta$ & $\alpha$ & $\theta_{s}^{b}$ & $k_{\rho}$ & $k_{\mu}$ \\
			\hline
			1 & Viscoelastic &0.0076  & 10 & 0.1 & 0 & $180^{o}$ & $\frac{1}{839}$  & $\frac{1}{58.8}$ \\
			\hline
			2 & Newtonian &0.0076  & 0 & 0 & 0 & $180^{o}$ & $\frac{1}{839}$  & $\frac{1}{58.8}$ \\
			\hline
			3 & Viscoelastic & 0.0373  & 10 & 0.1 & 0 & $180^{o}$ & $\frac{1}{839}$  & $\frac{1}{58.8}$ \\
			\hline
			4 & Newtonian & 0.0373  & 0 & 0 & 0 & $180^{o}$ & $\frac{1}{839}$  & $\frac{1}{58.8}$ \\
			\hline
		\end{tabular}
	\end{center}
\end{table}

\begin{table}[H]
	\begin{center}
		\caption{ Dimensionless numbers used in the simulations at large Ohnesorge numbers, see section 4.2.2}
		\label{Tabel 3}
		\begin{tabular}{c c c c c c c c c}
			\hline
			$Case$ & Droplets & $Oh$ & $Wi$ & $\beta$ & $\alpha$ & $\theta_{s}^{b}$ & $k_{\rho}$ & $k_{\mu}$ \\
			\hline
			1 & Viscoelastic & 0.118  & 10 & 0.1 & 0 & $180^{o}$ & $\frac{1}{839}$  & $\frac{1}{58.8}$ \\
			\hline
			2 & Newtonian & 0.118  & 0 & 0 & 0 & $180^{o}$ & $\frac{1}{839}$  & $\frac{1}{58.8}$ \\
			\hline
			3 & Viscoelastic & 0.1668  & 10 & 0.1 & 0 & $180^{o}$ & $\frac{1}{839}$  & $\frac{1}{58.8}$ \\
			\hline
			4 & Newtonian & 0.1668  & 0 & 0 & 0 & $180^{o}$ & $\frac{1}{839}$  & $\frac{1}{58.8}$ \\
			\hline
			5 & Viscoelastic & 0.3  & 10 & 0.1 & 0 & $180^{o}$ & $\frac{1}{839}$  & $\frac{1}{58.8}$ \\
			\hline
			6 & Newtonian & 0.3  & 0 & 0 & 0 & $180^{o}$ & $\frac{1}{839}$  & $\frac{1}{58.8}$ \\
			\hline
		\end{tabular}
	\end{center}
\end{table}

\begin{table}[H]
	\begin{center}
		\caption{
			Dimensionless numbers used in the simulations focusing on the role of shear thinning, see section 4.3}
		\label{Tabel 4}
		\begin{tabular}{c c c c c c c c c}
			\hline
			$Case$ & Droplets & $Oh$ & $Wi$ & $\beta$ & $\alpha$ & $\theta_{s}^{b}$ & $k_{\rho}$ & $k_{\mu}$ \\
			\hline
			1 & Viscoelastic & 0.118  & 10 & 0.1 & 0 & $180^{o}$ & $\frac{1}{839}$  & $\frac{1}{58.8}$ \\
			\hline
			2 & Viscoelastic & 0.118  & 10 & 0.1 & 0.2 & $180^{o}$ & $\frac{1}{839}$  & $\frac{1}{58.8}$ \\
			\hline
			3 & Viscoelastic & 0.118  & 10 & 0.1 & 0.5 & $180^{o}$ & $\frac{1}{839}$  & $\frac{1}{58.8}$ \\
			\hline
		\end{tabular}
	\end{center}
\end{table}

\begin{table}[H]
	\begin{center}
		\caption{Dimensionless numbers used in the simulations focusing on the role of the polymer viscosity ratio, see section 4.3}
		\label{Tabel 5}
		\begin{tabular}{c c c c c c c c c}
			\hline
			$Case$ & Droplets & $Oh$ & $Wi$ & $\beta$ & $\alpha$ & $\theta_{s}^{b}$ & $k_{\rho}$ & $k_{\mu}$ \\
			\hline
			1 & Newtonian & 0.118  & 0 & 0 & 0 & $180^{o}$ & $\frac{1}{839}$  & $\frac{1}{58.8}$ \\
			\hline
			2 & Viscoelastic & 0.118  & 10 & 0.1 & 0 & $180^{o}$ & $\frac{1}{839}$  & $\frac{1}{58.8}$ \\
			\hline
			3 & Viscoelastic & 0.118  & 10 & 0.5 & 0 & $180^{o}$ & $\frac{1}{839}$  & $\frac{1}{58.8}$ \\
			\hline
			4 & Viscoelastic & 0.118  & 10 & 0.6 & 0 & $180^{o}$ & $\frac{1}{839}$  & $\frac{1}{58.8}$ \\
			\hline
			5 & Viscoelastic & 0.118  & 10 & 0.7 & 0 & $180^{o}$ & $\frac{1}{839}$  & $\frac{1}{58.8}$ \\
			\hline
			6 & Viscoelastic & 0.118  & 10 & 0.8 & 0 & $180^{o}$ & $\frac{1}{839}$  & $\frac{1}{58.8}$ \\
			\hline
		\end{tabular}
	\end{center}
\end{table}

\bibliography{mybibfile}

\begin{thebibliography}{36}
\providecommand{\natexlab}[1]{#1}
\providecommand{\url}[1]{\texttt{#1}}
\expandafter\ifx\csname urlstyle\endcsname\relax
  \providecommand{\doi}[1]{doi: #1}\else
  \providecommand{\doi}{doi: \begingroup \urlstyle{rm}\Url}\fi

\bibitem[Abels et~al.(2012)Abels, Garcke, and Grün]{Abels2012}
H.~Abels, H.~Garcke, and G.~Grün.
\newblock Thermodynamically consistent, frame indifferent diffuse interface
  models for incompressible two-phase flows with different densities.
\newblock \emph{Math Models Methods Appl Sci.}, 22(03):\penalty0 1150013, 2012.

\bibitem[B.~Boreyko and Chen(2009)]{Bor2009}
J.~B.~Boreyko and C.-H. Chen.
\newblock Self-propelled dropwise condensate on superhydrophobic surfaces.
\newblock \emph{Phys. Rev. Lett.}, 103\penalty0 (2):\penalty0 184501, 2009.

\bibitem[Bazesefidpar et~al.(2021)Bazesefidpar, Brandt, and
  Tammisola]{Bazesefidpar2021}
K.~Bazesefidpar, L.~Brandt, and O.~Tammisola.
\newblock A dual resolution phase-field solver for wetting of viscoelastic
  droplets.
\newblock \emph{arXvi}, page 2111.04577v1, 2021.

\bibitem[Borges et~al.(2008)Borges, Carmona, Costa, and Don]{Borges2008}
R.~Borges, M.~Carmona, B.~Costa, and W.~Don.
\newblock An improved weighted essentially non-oscillatory scheme for
  hyperbolic conservation laws.
\newblock \emph{J. Comput. Phys.}, 227(6):\penalty0 3191--3211, 2008.

\bibitem[Boyaval et~al.(2009)Boyaval, Lelièvre, and Mangoubi]{Boyaval2009}
S.~Boyaval, T.~Lelièvre, and C.~Mangoubi.
\newblock Free-energy-dissipative schemes for the oldroyd-b model.
\newblock \emph{ESAIM: Mathematical Modelling and Numerical
  Analysis-Modélisation Mathématique et Analyse Numérique}, 43(3):\penalty0
  523--561, 2009.

\bibitem[Cha et~al.(2016)Cha, Xu, Sotelo, Chun, Yokoyama, Enright, and
  Miljkovic]{Cha2016}
H.~Cha, C.~Xu, J.~Sotelo, J.~Chun, Y.~Yokoyama, R.~Enright, and N.~Miljkovic.
\newblock Coalescence-induced nanodroplet jumping.
\newblock \emph{Phys. Rev. Fluids}, 1(6):\penalty0 064102, 2016.

\bibitem[Cheng et~al.(2018)Cheng, Xu, and Sui]{Xu2016}
Y.~Cheng, J.~Xu, and Y.~Sui.
\newblock Numerical investigation of coalescence-induced droplet jumping on
  superhydrophobic surfaces for efficient dropwise condensation heat transfer.
\newblock \emph{Int. J. Heat Mass Transf.}, 95:\penalty0 506--516, 2018.

\bibitem[Dekker et~al.(2022)Dekker, Hack, Tewes, Datt, Bouillant, and
  Snoeijer]{Dekker}
P.~Dekker, M.~Hack, W.~Tewes, C.~Datt, A.~Bouillant, and J.~Snoeijer.
\newblock When elasticity affects drop coalescence.
\newblock \emph{PRL}, 128(2):\penalty0 028004, 2022.

\bibitem[Dong(2012)]{Dong2012}
S.~Dong.
\newblock On imposing dynamic contact-angle boundary conditions for
  wall-bounded liquid–gas flows.
\newblock \emph{Comput Methods Appl Mech Eng.}, 247:\penalty0 179--200, 2012.

\bibitem[Enright et~al.(2014)Enright, Miljkovic, Alvarado, Kim, and
  Rose]{Enright2014}
R.~Enright, N.~Miljkovic, J.~Alvarado, K.~Kim, and J.~Rose.
\newblock Dropwise condensation on micro-and nanostructured surfaces.
\newblock \emph{Nanoscale Microscale Thermophys. Eng.}, 18:\penalty0 223--250,
  2014.

\bibitem[Farokhirad et~al.(2015)Farokhirad, Morris, and Lee]{Farokhirad2015}
S.~Farokhirad, J.~Morris, and T.~Lee.
\newblock Coalescence-induced jumping of droplet: Inertia and viscosity
  effects.
\newblock \emph{Phys. Fluids}, 27(10):\penalty0 102102, 2015.

\bibitem[Fattal and Kupferman(2004)]{Fattal2004}
R.~Fattal and R.~Kupferman.
\newblock Constitutive laws for the matrix-logarithm of the conformation
  tensor.
\newblock \emph{J Nonnewton Fluid Mech}, 123(2-3):\penalty0 281--285, 2004.

\bibitem[Fattal and Kupferman(2005)]{Fattal2005}
R.~Fattal and R.~Kupferman.
\newblock Time-dependent simulation of viscoelastic flows at high weissenberg
  number using the log-conformation representation.
\newblock \emph{J Nonnewton Fluid Mech}, 126(1):\penalty0 23--37, 2005.

\bibitem[Fuqiang et~al.(2020)Fuqiang, Li, Ni, and Wen]{Chu2020}
C.~Fuqiang, S.~Li, Z.~Ni, and D.~Wen.
\newblock Departure velocity of rolling droplet jumping.
\newblock \emph{Langmuir}, 31(14):\penalty0 3713--3719, 2020.

\bibitem[Gottlieb and Shu(1998)]{Gottlieb1998}
S.~Gottlieb and C.~Shu.
\newblock Total variation diminishing runge-kutta schemes.
\newblock \emph{Math. Comput.}, 67(221):\penalty0 73--85, 1998.

\bibitem[Jacqmin(2000)]{Jacqmin2000}
D.~Jacqmin.
\newblock Contact-line dynamics of a diffuse fluid interface.
\newblock \emph{J. Fluid Mech.}, 402:\penalty0 57--88, 2000.

\bibitem[Khismatullin and Nadim(2001)]{Khismatullin2001}
D.~Khismatullin and A.~Nadim.
\newblock Shape oscillations of a viscoelastic drop.
\newblock 63(6):\penalty0 061508, 2001.

\bibitem[Li et~al.(2020)Li, Chu, Zhang, Brutin, and Wen]{Li2020}
S.~Li, F.~Chu, J.~Zhang, D.~Brutin, and D.~Wen.
\newblock Droplet jumping induced by coalescence of a moving droplet and a
  static one: Effect of initial velocity.
\newblock \emph{Chem. Eng. Sci.}, 211:\penalty0 115252, 2020.

\bibitem[Liu et~al.(2014)Liu, Ghigliotti, Feng, and Chen]{Liu2014}
F.~Liu, G.~Ghigliotti, J.~Feng, and C.~Chen.
\newblock Numerical simulations of self-propelled jumping upon drop coalescence
  on non-wetting surfaces.
\newblock \emph{J. Fluid Mech.}, 752:\penalty0 39, 2014.

\bibitem[Lo et~al.(2014)Lo, Wang, and Lu]{Lo2014}
C.~Lo, C.~Wang, and M.~Lu.
\newblock Scale effect on dropwise condensation on superhydrophobic surfaces.
\newblock \emph{ACS Appl. Mater. Interfaces}, 6(16):\penalty0 14353--14359,
  2014.

\bibitem[Magaletti et~al.(2013)Magaletti, Picano, Chinappi, Marino, and
  Casciola]{Magaletti2013}
F.~Magaletti, F.~Picano, M.~Chinappi, L.~Marino, and C.~Casciola.
\newblock The sharp-interface limit of the cahn-hilliard/navier-stokes model
  for binary fluids.
\newblock \emph{J. Fluid Mech.}, 714:\penalty0 95, 2013.

\bibitem[Mokbel et~al.(2018)Mokbel, Abels, and Aland]{Mokbel2018}
D.~Mokbel, H.~Abels, and S.~Aland.
\newblock A phase-field model for fluid–structure interaction.
\newblock \emph{J. Comput. Phys.}, 372:\penalty0 823--840, 2018.

\bibitem[Mulroe et~al.(2017)Mulroe, Srijanto, Ahmadi, Collier, and
  Boreyko]{Mulroe2017}
M.~Mulroe, B.~Srijanto, S.~Ahmadi, C.~Collier, and J.~Boreyko.
\newblock Tuning superhydrophobic nanostructures to enhance jumping-droplet
  condensation.
\newblock \emph{ACS Appl. Mater. Interfaces}, 11(8):\penalty0 8499--8510, 2017.

\bibitem[Peng et~al.(2020)Peng, Yan, Li, Li, Cha, Ding, Dang, Jia, and
  Miljkovic]{Peng2020}
Q.~Peng, X.~Yan, J.~Li, L.~Li, H.~Cha, Y.~Ding, C.~Dang, L.~Jia, and
  N.~Miljkovic.
\newblock Breaking droplet jumping energy conversion limits with
  superhydrophobic microgrooves.
\newblock \emph{Langmuir}, 36:\penalty0 9510--9522, 2020.

\bibitem[Qian et~al.(2003)Qian, Wang, and Sheng]{Qian2003}
T.~Qian, X.~Wang, and P.~Sheng.
\newblock Molecular scale contact line hydrodynamics of immiscible flows.
\newblock \emph{Phys. Rev. E.}, 68(1):\penalty0 016306, 2003.

\bibitem[Wang et~al.(2016)Wang, Liang, Jiang, Zheng, Lan, and Ma]{Wang2016}
K.~Wang, Q.~Liang, R.~Jiang, Y.~Zheng, Z.~Lan, and X.~Ma.
\newblock Self-enhancement of droplet jumping velocity: the interaction of
  liquid bridge and surface texture.
\newblock \emph{RSC advances}, 101:\penalty0 99314--99321, 2016.

\bibitem[Wang and Ming(2019)]{Wang2019}
Y.~Wang and P.~Ming.
\newblock Dynamic and energy analysis of coalescence-induced self-propelled
  jumping of binary unequal-sized droplets.
\newblock \emph{Phys. Fluids}, 31(12):\penalty0 122108, 2019.

\bibitem[Wasserfall et~al.(2017)Wasserfall, Figueiredo, Kneer, Rohlfs, and
  Pischke]{Wasserfall2017}
J.~Wasserfall, P.~Figueiredo, R.~Kneer, W.~Rohlfs, and P.~Pischke.
\newblock Coalescence-induced droplet jumping on superhydrophobic surfaces:
  Effects of droplet mismatch.
\newblock \emph{Phys. Rev. Fluids}, 36(14):\penalty0 123601, 2017.

\bibitem[Watson et~al.(2015)Watson, Green, Schwarzkopf, Li, Cribb, Myhra, and
  Watson]{Watson2015}
G.~S. Watson, D.~W. Green, L.~Schwarzkopf, X.~Li, B.~W. Cribb, S.~Myhra, and
  J.~A. Watson.
\newblock A gecko skin micro/nano structure–a low adhesion, superhydrophobic,
  anti-wetting, self-cleaning, biocompatible, antibacterial surface.
\newblock \emph{Acta Biomater.}, 21:\penalty0 109--122, 2015.

\bibitem[Wisdom et~al.(2013)Wisdom, Watson, Qu, Liu, Watson, and
  Chen]{Wisdom2013}
K.~Wisdom, J.~Watson, X.~Qu, F.~Liu, G.~Watson, and C.~Chen.
\newblock Self-cleaning of superhydrophobic surfaces by self-propelled jumping
  condensate.
\newblock \emph{Proceedings of the National Academy of Sciences}, 110:\penalty0
  7992--7997, 2013.

\bibitem[Xu et~al.(2018)Xu, Di, and Yu]{Xu2018}
X.~Xu, Y.~Di, and H.~Yu.
\newblock Sharp-interface limits of a phase-field model with a generalized
  navier slip boundary condition for moving contact lines.
\newblock \emph{J. Fluid Mech.}, 849:\penalty0 805--833, 2018.

\bibitem[Yan et~al.(2019{\natexlab{a}})Yan, Zhang, Sett, Feng, Zhao, Huang,
  Vahabi, Kota, Chen, and Miljkovic]{Vahabi2017}
X.~Yan, L.~Zhang, S.~Sett, L.~Feng, C.~Zhao, Z.~Huang, H.~Vahabi, A.~Kota,
  F.~Chen, and N.~Miljkovic.
\newblock Droplet jumping: effects of droplet size, surface structure, pinning,
  and liquid properties.
\newblock \emph{ACS nano}, 13(2):\penalty0 1309--1323, 2019{\natexlab{a}}.

\bibitem[Yan et~al.(2019{\natexlab{b}})Yan, Zhang, Sett, Feng, Zhao, Huang,
  Vahabi, Kota, Chen, and Miljkovic]{Yan2019}
X.~Yan, L.~Zhang, S.~Sett, L.~Feng, C.~Zhao, Z.~Huang, H.~Vahabi, A.~Kota,
  F.~Chen, and N.~Miljkovic.
\newblock Droplet jumping: effects of droplet size, surface structure, pinning,
  and liquid properties.
\newblock \emph{ACS nano}, 13(2):\penalty0 1309--1323, 2019{\natexlab{b}}.

\bibitem[Yuan et~al.(2019)Yuan, Hu, Chu, and Wu]{Yuan2019}
Z.~Yuan, Z.~Hu, F.~Chu, and X.~Wu.
\newblock Enhanced and guided self-propelled jumping on the superhydrophobic
  surfaces with macrotexture.
\newblock \emph{Appl. Phys}, 115:\penalty0 163701, 2019.

\bibitem[Yue et~al.(2004)Yue, Feng, Liu, and Shen]{Yue2004}
P.~Yue, J.~Feng, C.~Liu, and J.~Shen.
\newblock A diffuse-interface method for simulating two-phase flows of complex
  fluids.
\newblock \emph{J. Fluid Mech.}, 515:\penalty0 293, 2004.

\bibitem[Zhang et~al.(2013)Zhang, He, Chen, Wang, Song, and Jiang]{Zhang2013}
Q.~Zhang, M.~He, J.~Chen, J.~Wang, Y.~Song, and L.~Jiang.
\newblock Anti-icing surfaces based on enhanced self-propelled jumping of
  condensed water microdroplets.
\newblock \emph{ChemComm}, 49:\penalty0 4516--4518, 2013.

\end{thebibliography}

\end{document}